\documentclass[journal=aamick,manuscript=letter,layout=twocolumn]{achemso}

\usepackage[version=3]{mhchem} 
\usepackage{graphicx}
\usepackage{dcolumn}
\usepackage{bm}
\usepackage{color,times}
\usepackage[mathlines]{lineno}
\usepackage{amssymb}
\usepackage{amsmath}
\usepackage{natbib}
\usepackage{soul,color}

\title{High thermoelectric performance in two-dimensional
tellurium: An \textit{ab initio} study}

\author{Zhibin Gao}
\affiliation{Center for Phononics and Thermal Energy Science,
             China-EU Joint Center for Nanophononics,
             Shanghai Key Laboratory of Special Artificial
             Microstructure Materials and Technology,
             School of Physics Sciences and Engineering,
             Tongji University, Shanghai 200092, China}

\author{Gang Liu}
\email
            {liugang8105@gmail.com}
\affiliation{School of Physics and Engineering, Henan
             University of Science and Technology,
             Luoyang 471023, China}

\author{Jie Ren}
\email
            {Xonics@tongji.edu.cn}%
\affiliation{Center for Phononics and Thermal Energy Science,
             China-EU Joint Center for Nanophononics,
             Shanghai Key Laboratory of Special Artificial
             Microstructure Materials and Technology,
             School of Physics Sciences and Engineering,
             Tongji University, Shanghai 200092, China}

\date{\today} 


\begin{document}

\begin{abstract}
In 2016, bulk tellurium was experimentally observed as a remarkable thermoelectric material.
%
Recently, two-dimensional (2D) tellurium, called tellurene,
has been synthesized and has exhibited unexpected electronic properties compared with
the 2D MoS$_2$. They have also
been fabricated into air-stable and high efficient field-effect
transistors. There are two stable 2D tellurene
phases. One ($\beta$-Te) has been confirmed with an ultralow lattice thermal conductivity ($\kappa_L$).
However, the study of the transport properties of the other more stable phase,
$\alpha$-Te, is still lacking. Here, we report the thermoelectric
performance and phonon properties of $\alpha$-Te using Boltzmann transport
theory and first principle calculations. A maximum \textit{ZT} value of 0.83
is achieved under reasonable hole concentration,
suggesting that the monolayer $\alpha$-Te is a potential competitor in the thermoelectric field.
\end{abstract}

\section{Introduction}
Recently, 2D materials have triggered a large number of interest
due to the striking physical properties related to the low
dimensionality~\cite{novoselov2004,mounet2018,dresselhaus2007}.
Graphene~\cite{novoselov2004} in group-IV, borophene in group-III~\cite{tang2007,li2018stretch}
and black phosphorous in group-V~\cite{guan2014,li2014black} have been extensively
investigated experimentally and theoretically. However, researches related to group-VI
monolayer materials are still lacking to date.

Recently, Zhu \textit{et al.}~\cite{zhu2017} firstly have proposed three
novel types of monolayer tellurene, namely $\alpha$-, $\beta$- and $\gamma$-Te.
Afterwards, tellurene has been synthesized and attracted lots of
interest~\cite{zhu2017,huang2017,wang2018,du2017,gao2018,chen2017ultrathin,liu2018,qiao2018}.
These findings indicate that tellurene possesses unusually physical properties,
especially for applications in electronic and thermoelectric devices. Furthermore, Wang \textit{et al.}~\cite{wang2018} pointed out that
tellurene can be obtained in a very large scale and can be fabricated
into air-stable and high-performance field-effect transistors, which makes it
different from %
other 2D materials. Furthermore, tellurene has much
larger carrier mobility than that of 2D MoS$_2$~\cite{zhu2017}. Few layer tellurene
also has extraordinarily electronic transport properties~\cite{wang2018,liu2018,qiao2018}.

$\beta$-Te has the intrinsic $\kappa_L$ of 2.16 and 4.08~W/mK along both
directions (300 K)~\cite{gao2018} and also shows significant
anisotropy, indicating good potential in application of thermoelectric devices with
\textit{ZT} = 0.8~\cite{sharma2018}. Besides, in 2016, bulk tellurium was
experimentally observed as a remarkable thermoelectric material~\cite{lin2016}.
Energetically, monolayer $\alpha$-Te is the most stable structure compared with
the metastable $\beta$- and $\gamma$-Te phases~\cite{zhu2017}. As the most stable
phase, $\alpha$-Te is also a promising one to be synthesized experimentally and
has applications in practical nanoelectronics.

Generally, we use \textit{ZT} to assess the degree of thermoelectric material by
$ZT={S^2\sigma T/(\kappa_e+\kappa_L)}$, where \textit{S} means Seebeck coefficient,
$\sigma$ is the electric conductivity, \textit{T} is the absolute temperature,
$\kappa_e$ and $\kappa_L$ are the electronic and lattice thermal conductivity, respectively.
Ideally, one would like to enhance the numerator and weaken the denominator at the same time.
Fortunately, one can enhance the scattering of phonons by boundary, defect, isotopic
effect, resonant bonding~\cite{lee2014resonant}, rattle-like scattering~\cite{tadano2015impact},
lone electron pairs~\cite{jana2016origin}, making composite structures~\cite{snyder2008}
and even forming solid solutions~\cite{liu2012copper}. Unfortunately, $S$ and $\sigma$ unusually
interweave and behave in an opposite trend, which induces the complexity to efficiently realize the
waste heat recovery~\cite{snyder2008}.

Interestingly, low dimensional such as 2D materials have the potential to break above relation
between S and $\sigma$ due to the quantum confinement effect pointed by Dresselhaus \textit{et al.}~\cite{dresselhaus2007}.
%
%
There are three phases of atomically thin tellurium, namely $\alpha$-, $\beta$- and $\gamma$-Te.
The first two are energetically, dynamically and mechanically stable.
Furthermore, $\alpha$-Te is more stable than $\beta$-Te~\cite{zhu2017}
that has an unusually low $\kappa_L$~\cite{gao2018} and superior
thermoelectric property~\cite{sharma2018}. However, relevant works
about $\kappa_L$ and thermoelectric properties of the more stable
$\alpha$-Te phase is still lacking. In this letter, we focus on the
more stable phase, $\alpha$-Te, and would like to explore its novel
electronic and thermal properties. We find a largest
\textit{ZT} (0.83) is achieved under reasonable hole concentration at
700~K for $\alpha$-Te.

\section{Computational methods}
We use vasp code and PBE functional~\cite{PAW1,PAW2,PBE1996}. The cutoff
energy is 500~eV. In order to obtain a more accurate electronic band
structure and density of states, we use HSE06 approach~\cite{HSE06}.
The van der Waals (vdW) correction proposed by Grimme~\cite{grimme2006}
was taken into consideration in our calculations. The criterion of
convergence energy and force are 10$^{-6}$ eV and 1~meV/{\AA}.
We use 17$\times$17$\times$1 k-mesh in the Brillouin zone.

In the framework of Boltzmann transport theory (BTE),
the electrical transport properties, such as $\sigma$, \textit{S}
and $\kappa{_e}$ can be expressed as~\cite{madsen2006,chen2013electronic}:
\begin{equation} %
\begin{aligned}
\label{tensor}
\textbf{K$_n$}=&\frac{1} {4 \pi^3} \sum_{i,\textbf{k}}\tau_i(\textbf{k}) \textbf{v}_i(\textbf{k})
\otimes \textbf{v}_i(\textbf{k}) [\varepsilon_i(\textbf{k}) - \mu ]^n\\
&\times [-\frac{\partial f(\mu, T, \varepsilon_i)} {\partial \varepsilon_i}],
\end{aligned}
\end{equation}
\begin{equation} %
\label{sigma}
\sigma = e^2 \textbf{K$_0$},
\end{equation}
\begin{equation} %
\label{seebeck}
S = \frac{1} {e T} \textbf{K$_1$} \textbf{K$_0$}^{-1},
\end{equation}
\begin{equation} %
\label{electronic_kaapa}
\textit{k$_e$} =  \frac{1} {T} (\textbf{K}_2 - \textbf{K}_1^2 \textbf{K}_0^{-1}). \\
\end{equation}
where $\tau_i(\textbf{k})$ and $\varepsilon_i$ are the electronic relaxation time and
energy eigenvalue. \textit{k$_e$} can be obtained based on the tensor \textbf{K$_n$}.

The in-plane $\kappa_L$ under relaxation time approximation can be obtained~\cite{gao2018}:
\begin{equation} %
\label{kappa}
\kappa_{\alpha\beta} = \frac{1} {V} \sum_{\lambda} C_\lambda
                       \upsilon_{\lambda \alpha}
                       \upsilon_{\lambda \beta} \tau_{\lambda},  \\
\end{equation}
where the meaning of each parameter are explained elsewhere~\cite{gao2018}.
We use Phonopy~\cite{phonopy2008} and ShengBTE~\cite{ShengBTE2014} to deal
with the harmonic and anharmonic force constants.
We use the equation (4) to calculate the electronic
thermal conductivity \textit{k$_e$} since sometimes the results
from the Wiedemann-Franz law is unreasonable, such as in metallic
VO$_2$~\cite{lee2017anomalously}, clean graphene near the charge
neutral point~\cite{crossno2016observation}, and single-electron
transistor~\cite{dutta2017thermal}. As a matter of fact, the
Wiedemann–Franz law is only suitable for the system where the
scattering of electrons in the material is dominanted by the elastic
collision (``good metal'')~\cite{mosso2017heat}. Therefore, we use
the definition to calculate \textit{k$_e$}. It should be noted that
an effective thickness should be defined in order to calculate the
electronic and thermal properties for 2D materials. The thickness of
$\alpha$-Te is 7.74~{\AA}, whose definition is clearly defined in
elsewhere~\cite{gao2017,gao2018}.

\section{Results and discussion}
%
The optimized structure of
$\alpha$-Te is shown in Figure 1a and 1b with lattice constant
4.238~{\AA} obtained by PBE+D2 functional. The $\alpha$-Te possesses
\textit{P}-3M1 (164) symmetry group belonging to trigonal system and
isotropic pattern in 2D plane, which is quite different from the
anisotropic bulk Te~\cite{peng2014} and 2D $\beta$-Te~\cite{zhu2017,gao2018}.
From the top and side views, $\alpha$-Te and sandwiched 1T-MoS$_2$
look alike in many ways. Intermediate Te atom is octahedrally
coordinated to six neighboring Te atoms and upper and lower Te
layers form the A-B like stacking. The coordination numbers in
outer and centered Te are 3 and 6, which is the characteristic
of a multivalency formation of Te atom located in the near bottom
of the periodic table. The bond length in $\alpha$-Te is all
3.04~{\AA}, which is larger than those of 2.77~{\AA} and
3.03~{\AA} (two type bonds) in monolayer $\beta$-Te~\cite{gao2018}.

Bulk Te was reported a direct semiconductor (E$_g$ = 0.25~eV)~\cite{peng2014}
and it also has recently been experimentally observed as a remarkable
thermoelectric material~\cite{lin2016}.
Figure 1c and 1d show the electronic band structure and density of
states (DOS) of $\alpha$-Te, indicating a near-direct band gap
material (E$_g$ = 1.11~eV) at HSE06 level. Our result is very
consistent with the previous theoretical work~\cite{zhu2017}.

A good thermoelectric material not only needs a minimum thermal conductivity, but also requires a
simultaneously maximum power factor (\textit{PF}). However,
Seebeck \textit{S} and $\sigma$ are usually inter-weaved.
A large \textit{S} requires a large
carrier effective mass decided by~\cite{heremans2008}
$ S = \frac{8 \pi^2 k_B^2 T } {3eh^2} m^* (\frac{ \pi } {3n})^{2/3} $
in 3D semiconducting materials,
but $\sigma$ is inversely proportional to the carrier effective
mass due to $\mu = \frac{e \tau } {m^*}$.%
In 1993, Hicks and Dresselhaus proposed a seminal idea that sharpen the energy
dependence of the DOS in 1D and 2D systems to alleviate the coupling between \textit{S} and $\sigma$,
consequently enhancing the \textit{PF}~\cite{hicks1993effect,hicks1993thermoelectric}. Mahan and Sofo
further generalized it to a refined sentence~\cite{mahan1996best}. Afterwards, these guiding
principles triggered two interesting band structure shapes. One is
``pudding-mold''~\cite{kuroki2007pudding,usui2013large} and the other one is
``Mexican-hat-shape''~\cite{ge2017first}. Recently, a backward thinking emerges, which
introduce quasi-one-dimensional electronic band dispersions in 2D or 3D materials
to increase the thermoelectric performance~\cite{mi2015enhancing,bilc2015low}.
A relatively flat band means a large DOS in a narrow energy region near the Fermi
level. The dispersive band leads to a high carrier velocity, also indicates a small $m^*$
and therefore a high $\mu$~\cite{yang2012power,yang2011trends,yang2008evaluation,yang2016tuning}.
Interestingly, The valence bands near the Fermi level of $\alpha$-Te
shown in Figure 1c has some hole pockets and relatively flat bands,
which is the characteristic of a good thermoelectric property and
is similar to renowned PbTe$_{1-x}$Se$_x$ thermoelectrics~\cite{pei2011}.

Based on the band structure, we could evaluate the \textit{S}, $\sigma$,
and $\tau$ of $\alpha$-Te. Since the melting point of tellurium is 723~K,
we select three typical temperatures (300, 500 and 700 K) in the whole
calculations. Figure 2a shows the Seebeck coefficients \textit{S} in the
n- and p-type of $\alpha$-Te as functions of carrier concentration.
Overall, the absolute values of \textit{S} decrease for both n- and p-type
when increasing the carrier concentration, which reflects the inverse
proportion between \textit{S} and carrier concentration \textit{n}. Moreover,
the absolute value of \textit{S} of hole doping is bigger than the electron
doping in $\alpha$-Te. For instance, the value of $|\textit{S}|$ for n-type is
242.9~$\mu$V/K at 10$^{12}$ cm$^{-2}$ carrier concentration at 500~K
temperature, only around half of p-type 529.6~$\mu$V/K at same condition.
The maximum value \textit{S} of p-type $\alpha$-Te is around 700~$\mu$V/K at room
temperature, which is the double of that 350~$\mu$V/K in bulk tellurium~\cite{lin2016}.
This is the physical reason that the \textit{ZT} of $\alpha$-Te are comparable
with bulk tellurium, although the $\kappa_L$ in $\alpha$-Te (9.85 W/mK) is
around 7 times of that in bulk tellurium (1.5 W/mK)~\cite{lin2016}. In this
sense, if one further increases the phonon scattering using some defects and
isotope effect in $\alpha$-Te, the $\kappa_L$ will decrease but with the same
level of Seebeck coefficient \textit{S}. Then the \textit{ZT} of $\alpha$-Te
will be significantly enhanced and surpass the bulk tellurium~\cite{lin2016}.

The difference between electron and hole doping in Seebeck \textit{S}
originates from the electronic band structure and DOS in Figure 1c and 1d. As we
discussed above, \textit{S} is primarily regulated by the effect mass m$^*$
and the magnitude of DOS. A more smooth valence band (hole-doping)
compared with a conduction band (electron-doping) corresponds to a larger
carrier m$^*$, indicating a large \textit{S} since \textit{S} is
proportional to the m$^*$. Furthermore, DOS of
p-type is obviously much larger than n-type of $\alpha$-Te around
the Fermi level shown in Figure 1d. These evidences verify that
the superior performance of the hole doping compared with the
electron doping in $\alpha$-Te.

According to the BTE, the electronic conductivity $\sigma$ is dependent
and is proportional to the relaxation time $\tau$. Therefore, a reasonable
relaxation time $\tau$ should
be chosen. As a matter of fact, $\tau$ in materials is a function of
temperature and carrier concentration. So far, experimental measurement
may be the only effective way to solve this issue. For theoretical
calculations, $\tau$ is difficult to obtain with high precision.
Specifically, there are mainly three methods, such as using a
designated $\tau$ based on revelent experiments~\cite{bilc2015,he2016},
electro-phonon coupling implemented in EPW~\cite{ponce2016}. Another
approach is based on the well-known model: $\tau = \frac{m^* \mu } {e} $
in which $\mu$ is the carrier mobility and in 2D materials using
deformation potential theory acoustic phonon limited $\mu$ could be
expressed as~\cite{qiao2014}:
\begin{equation} %
\label{mobility}
\mu_{2D} = \frac{e \hbar^3 C_{2D} } {k_B T m^* m_d E_i^2 } \\
\end{equation}
where m$_d$ is the average effective mass dominated by
m$_d$ = $\sqrt{m_x^* m_y^*}$.
\textit{E$_i$} can be expressed
as $E{_i} = \Delta E{_i}/( \Delta l/l{_0} ) $ in which
$\Delta E{_i}$ is the \textit{i}$^{th}$ band under small compression
and expansion compared with the energy of unstrained system. Here, it
supposes that the energy of core electrons, considered as the energy
reference, do not change under the small deformation. This treating
process is reasonable and have also been verified by the previous
researches~\cite{qiao2014,mi2015enhancing}. The detailed calculations
is shown in the Supporting Information.
Our calculated effective masses $m^*$ are 0.107 $m_e$ for the conduction band
minimum and
0.164 $m_e$ for the valence band maximum and the carrier mobilities are
2.086$\times$10$^3$ cm$^2$V$^{-1}$ s$^{-1}$ for electrons and
1.736$\times$10$^3$ cm$^2$V$^{-1}$ s$^{-1}$ for holes.
Our result is very consistent with the previous report~\cite{zhu2017}.
Therefore, based
on $\tau = \frac{m^* \mu } {e} $, we can obtain the relaxation time
of 0.127~ps for electrons and 0.163~ps for holes in $\alpha$-Te.

%
Contrary to the Seebeck \textit{S}, electronic conductivity $\sigma$
increases with increasing carrier concentration. Figure 2b shows a
larger $\sigma$ of electron doping compared with hole doping at
the same carrier concentration due to the smaller effective
mass (0.107 m$_e$ for electrons $ < $ 0.164 m$_e$ for holes).
Furthermore, $\sigma$ of two type $\alpha$-Te are insensitive to
temperature, which is similar to that of $\beta$-Te~\cite{sharma2018}.

A high \textit{PF} needs a large \textit{S} and a large
$\sigma$ simultaneously. One should compromise between decreasing
function of \textit{S} and increasing function of $\sigma$.
The calculated power factor ($\textit{S}^2\sigma$)
that can demonstrate the thermoelectric performance of $\alpha$-Te is
shown in Figure 2c. The \textit{PF} for both types first increase, climb
the summit at moderate carrier concentrations, and then decrease when
increasing the concentration. The p-type \textit{PF} is significantly
larger than n-type of $\alpha$-Te, though they exhibit similar trend
dependent on the carrier concentration. For instance, the maximum
\textit{PF} of holes and electrons are 74.6~mW/mK$^2$ at
6.12$\times$10$^{13}$ concentration and 21.2~mW/mK$^2$ at
2.86$\times$10$^{13}$ concentration at 500~K. The former one is around
2.5~times larger than that of the latter one and
is 2~times larger than 2D SnSe~\cite{wang2015}.


%
More interestingly, the maximum \textit{PF} of bulk tellurium at room
temperature is around 1.3~mW/mK$^2$~\cite{lin2016}, which is around
a tenth of that in $\alpha$-Te (12~mW/mK$^2$) at the similar condition shown
in Figure 2c. The physical reason behind it is the very small effective
mass $m^*$ of $\alpha$-Te. The $m^*$ of standard 2H-MoS$_2$ is 0.47 $m_e$
for electrons and 0.58 $m_e$ for holes~\cite{zhu2017}. However, in
$\alpha$-Te, the $m^*$ is 0.107 m$_e$ for electron and  0.164 $m_e$ for
holes. Obviously, the effective mass of carriers in $\alpha$-Te is quite
smaller than 2H-MoS$_2$. A smaller $m^*$ will result in a larger electronic
conductivity qualitatively decided by the simple model:
$\sigma= \frac{ne^2\tau} {m^*}$. Moreover, Liu \textit{et al.}~\cite{liu2018}
pointed out that lone-pair electrons in bulk Te will significantly enhance the
interchain electronic transport. Similarly, the coordination number in $\alpha$-Te
is 3 for outer Te atoms, which is only half of valence electrons in Te atoms (group-VI).
Therefore, each outside Te atom exists 3 lone-pair electrons.
These lone-pair electrons will enhance the hopping term between electrons and
contribute to the superior electronic transport properties in
$\alpha$-Te~\cite{zhu2017,liu2018}.

An optimal \textit{ZT} also needs a minimum thermal conductivity
($\kappa_e+\kappa_L$). For electronic thermal conductivity,
\textit{k}$_e$ has been discussed in the computational methods. Next,
we would like to explore the phonon transport properties
of $\alpha$-Te, as phonon dominates the thermal transport in semiconductors.
The calculated phonon dispersions are shown in Figure 3a.
Firstly, it shows that all phonon branches are free from negative and
this result confirms the dynamical stability of $\alpha$-Te.
Moreover, there is no phonon gap between acoustic
and optical branches, indicating a strong optical-acoustic phonon scattering
which will suppress the $\kappa_L$~\cite{lindsay2013} of materials.

A high-order ($\geq$ 3) scattering of phonons leads to a limited
$\kappa_L$~\cite{gao2016heat,gao2016stretch}. Figure 3b shows the intrinsic
$\kappa_L$ computed with both non-iterative and iterative method
from 100~K to 700~K. We note that both
methods give similar trends of $\kappa_L$ in the whole temperature range. However,
the iterative method always gives values of $\kappa_L$ higher than
the non-iterative method at the same temperature. We choose the values
of iterative method in the following discussion. We find $\kappa_L$
of $\alpha$-Te can be well described by the \textit{T$^{-1}$} curve.
This phenomenon is also found in bulk Te~\cite{lin2016} and
$\beta$-Te~\cite{gao2018}. The value of intrinsic $\kappa_L$
is 9.85~W~m$^{-1}$K$^{-1}$ (300~K), which is twice as much as
that of $\beta$-Te at the same temperature~\cite{gao2018}.
Furthermore, $\kappa_L$ of $\alpha$-Te is also larger than bulk
Te (2.77~W/mK)~\cite{gao2018}, which is consistent with the familiar
trend ($\kappa_{2D}$ $>$ $\kappa_{3D}$)~\cite{seol2010two,balandin2011thermal,gao2018}.

The normalized contribution of ZA, TA, LA and optical branches varying
with temperature are plotted in Figure 3c, in which each mode varies small
at different temperatures especially beyond 200~K, and can be regarded as
independent of temperature, similar with the $\beta$-Te~\cite{gao2018}. At
room temperature, three acoustic branches contribute about 90\% to the
total intrinsic $\kappa_L$, while the summation of all optical phonons is
only 10\%. Furthermore, the proportions of
contributions to total $\kappa_L$ are about 25\%, 47\%, and 17\% for LA,
TA, and ZA phonons, in which ZA mode is the smallest
one among three acoustic branches in $\alpha$-Te, quite different from
graphene where the proportion of ZA mode is about 75\%~\cite{lindsay2014}.

It was reported that there is a symmetry selection rule in flat graphene that
only even number of ZA modes can be involved in scattering processes~\cite{lindsay2010,peng2016phonon}.
This rule originates from the reflection symmetry about c axis in the structure.
The thermal resistance roots in the high order of phonon scattering. Due
to the selection rule, ZA mode contributes mainly to the $\kappa$ in graphene.
However, the lattice does not have the reflection symmetry about c axis in
$\alpha$-Te. As a result the selection rule is broken, the importance of
ZA mode in thermal transport of $\alpha$-Te is much smaller than graphene.

%
%

Moreover, it is worthwhile to explore the contribution of phonons with
different frequencies to the total $\kappa_L$. Here the frequency-resolved
$\kappa_L$ for $\alpha$-Te at 300~K is displayed in Figure 3d. It can be found
that the main contribution originates from acoustic phonons with frequencies
lower than 2~THz, while the contribution of optical phonons is very small.
This is consistent with Figure 3c. Furthermore, the acoustic phonons with very
low frequency (lower than 1~THz) contribute most to $\kappa_L$. In the
following, we will find the behind reason is that phonons around $\Gamma$
point usually possess large phonon relaxation time.

All samples of material have finite size in practical experiment and device
application, thus the additional boundary scattering will reduce $\kappa_L$,
especially at nanoscale or at low temperatures. Usually, in nanostructures,
$\kappa_L$ related to boundary scattering can be evaluated as~\cite{balandin2011thermal,gao2018}:
\begin{equation}\label{BDR}
\frac{1} {\tau^B_\lambda} = \frac{\upsilon_\lambda} {L},  \\
\end{equation}
where \textit{L} is the size of material. This formula represents the situation
for a completely diffusive boundary scattering of phonons. Cumulative
$\kappa_L$ ($\wedge$ $<$ $\wedge_{max}$) can also be used to study the size
effect, whose definition has been defined elsewhere~\cite{ShengBTE2014}.
Figure 3e shows
the size dependence of $\kappa_L$ for $\alpha$-Te with completely diffusive boundary
scattering at room temperature and the cumulative $\kappa_L$ is displayed in
Figure 3f. In Figure 3e, $\kappa_L$ becomes lower while the sample size getting
smaller, as the boundary scattering becomes larger. For a specifical example,
when the sample size is 1~$\mu$m, $\kappa_L$ can be suppressed to
7.56~W~m$^{-1}$K$^{-1}$, about 78\% of value for the infinitely large system.
When the sample size is 10~$\mu$m, $\kappa_L$ reaches 91\% of the infinitely
large system. It implies the size effect always keeps till about 10~$\mu$~m,
such as 2D WTe$_2$~\cite{ma2016}.

The size effect can also be estimated from the cumulative $\kappa_L$ as a
function of MFP shown in Figure 3f. We can find that the MFP range of
phonons that contribute significantly to $\kappa_L$ is about 10~nm to
10~$\mu$m, which is consistent with our previous result in Figure 3e.
In order to obtain the representative MFP, we fit cumulative $\kappa_L$
using a single parametric function:
$\kappa(\wedge\leq\wedge_{max}) = \kappa_{max}/(1+\wedge_0/\wedge_{max})$,
where $\kappa_{max}$ is $\kappa_L$ of infinite size, $\wedge_{max}$ is the
maximal MFP. The fitting parameter $\wedge_0$ can be interpreted as the
typical MFP with which the phonons contribute significantly to $\kappa_L$.
The fitted curve is shown in Figure 3f by dashed line. The parameter
$\wedge_0$ is 94~nm, which could be interpreted as the representative MFP of $\alpha$-Te.
This implies the $\kappa_L$ of $\alpha$-Te may decline sharply with the sample size
of hundreds of nm. This property is propitious to the electronic and thermoelectric
materials based on $\alpha$-Te in the nanoscale~\cite{wang2018}.

Group velocity of phonon has great effects on thermal transport of materials,
as it is proportional to $\upsilon^2$ based on Eq. (3). The calculated $\upsilon^2$
with different frequencies are shown in Figure 4a. LA branch possesses the
highest $\upsilon$ among three acoustic branches, specifically more than
two times of that for ZA and TA phonons. The optical phonons around 3.7~THz
has the largest $\upsilon$ but they only contribute 10\% to total $\kappa_L$.
Here $\tau$ is also displayed in Figure 4b. $\tau$ around $\Gamma$ point has
the magnitude of 10$^3$-10$^4$~ps but $\tau$ of most phonons with frequency
more than 5~THz is about 0.1-10~ps. That is the reason why phonons with very
low frequency dominate the thermal transport in $\alpha$-Te (Figure 3d).
Furthermore, $\tau$ of TA branch is the largest of all, about 10$^4$~ps around
the $\Gamma$ point. That is the reason why TA mode contributes most to
$\kappa_L$ even more than LA phonons (Figure 3c). It can also explain that
optical phonons contribute little to $\kappa_L$ since most of them have
ultralow $\tau$ of only 1 to 10~ps.

Gr\"{u}nerisen parameter provides crucial information on the anharmonic
interactions of phonons. $\gamma$ in $\alpha$-Te
for whole frequency spectrum is displayed in Figure 4c.
$\gamma$ of TA and LA modes are in the range of 0-2, while those of optical modes are
1-6. Comparing with $\beta$-Te, $\gamma$ of $\alpha$-Te are smaller,
indicating a lower anharmonicity and a higher $\kappa_L$~\cite{gao2018}.
Interestingly, it can be found $\gamma$ of acoustic and optical phonon modes
have suddenly dropped around 2-2.5~THz. Since there is no acoustic-optical phonon gap, the
large acoustic-optical phonon scattering would lead to a giant $\gamma$
and is also the source of ultralow $\kappa_L$ in $\beta$-Te~\cite{gao2018}.
In the three-phonon scattering process, all phonon modes must be satisfied
the conservation of energy and momentum simultaneously~\cite{lindsay2008three}.
The quantitative probability of all available three-phonon scattering in
$\alpha$-Te are shown in Figure 4d. This \textit{P}$_3$ phase space is one
order smaller than of $\beta$-Te, indicating a weaker anharmonicity of
$\alpha$-Te compared with $\beta$-Te~\cite{gao2018}.

Based on the above results, intrinsic figure of
merit \textit{ZT} can be evaluated and exhibited in Figure 5. The climax
\textit{ZT} at 700~K are 0.57 and 0.83 for electron and hole doping
$\alpha$-Te, which is as high as $\beta$-Te~\cite{sharma2018} and has
the same order of bulk Te~\cite{lin2016}. Room temperature \textit{ZT}
of n-type $\alpha$-Te is 0.54.
Our calculation suggests that monolayer $\alpha$-Te is a promising thermoelectric
material.
Furthermore, it can be promoted further via suppressing $\kappa_L$
such as grain boundary, isotope effect and vacancy~\cite{lv2015experimental}.


\section{Conclusion}

In summary, we have systematically studied the extraordinarily high thermoelectric
performance of monolayer $\alpha$-Te by the \textit{ab initio} calculations.
To explain the high \textit{ZT}, we consider the electronic and thermal transport
properties.
A large and asymmetric electronic density of states induce relatively flat
bands, leading to a large Seebeck coefficient \textit{S}. A small effective
carrier mass $m^*$ of CBM (VBM) will result in a larger electronic conductivity
$\sigma$. Combining both, the \textit{PF} in $\alpha$-Te (12~mW/mK$^2$) exhibits
ten times as large as that in bulk tellurium (1.3~mW/mK$^2$)~\cite{lin2016} at
the similar condition.
For lattice thermal conductivity $\kappa_L$, $\alpha$-Te consists
of heavy atomic mass, which breeds low harmonic properties, such as low group
velocities. For anharmonicity, strong phonon anharmonic scattering rates including
acoustic-optical phonon scattering bring about a same order of magnitude
$\kappa_L$ of $\alpha$-Te compared with the bulk Te and $\beta$-Te.
A superior electronic properties and a suitable $\kappa_L$ combine together
leading to the ultrahigh \textit{ZT} in $\alpha$-Te. A maximum \textit{ZT}
of 0.83 is achieved with reasonable hole concentration at 700~K. Our results
indicate that monolayer $\alpha$-Te is a promising thermoelectric material.
We hope that $\alpha$-Te, more stable than $\beta$-Te, would be experimentally
observed with high \textit{ZT} in the near future.\\


\section{Supporting Information}
The Supporting Information is available free of charge on the
ACS Publications website via the Internet at
https://pubs.acs.org/journal/aamick.

Electronic band structure and effective mass around
conduction band minimum and valence band maximum of
two-dimensional tellurium.\\

{\noindent\bf Author Information}\\

{\noindent\bf Corresponding Author}\\
$^*$E-mail: {\tt liugang8105@gmail.com}\\
$^*$E-mail: {\tt Xonics@tongji.edu.cn}

{\noindent\bf ORCID}\\
Zhibin Gao: 0000-0002-6843-381X \\
Jie Ren: 0000-0003-2806-7226

{\noindent\bf Notes}\\
The authors declare no competing financial interest.



\begin{acknowledgement}
We thank our member Yi Wang for helpful discussions.
This work is supported by the National Natural Science
Foundation of China (No. 11775159), the Natural Science
Foundation of Shanghai (No. 18ZR1442800), the National
Youth 1000 Talents Program in China, and the Opening
Project of Shanghai Key Laboratory of Special Artificial
Microstructure Materials and Technology. Computational
resources have been provided by the Tongji University.

\end{acknowledgement}

\newpage

\begin{figure*}[tbp]
\begin{center}
\includegraphics[width=5.0in]{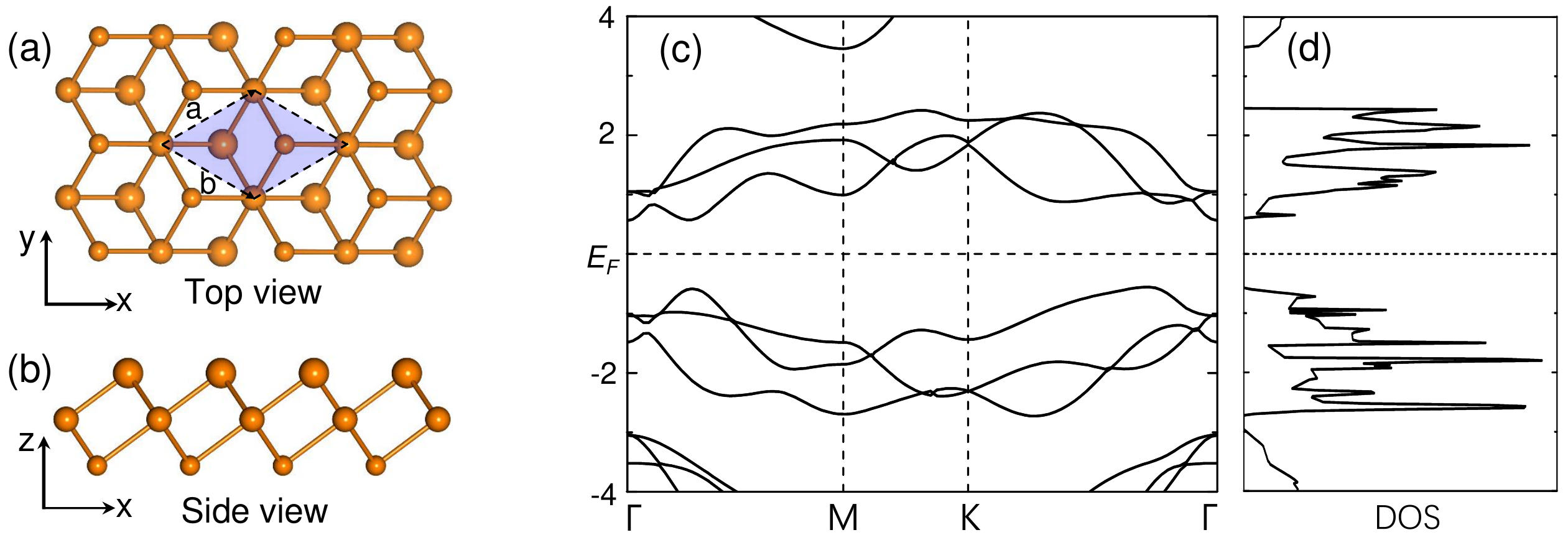}
\end{center}
\caption{(a)(b) Top and side views of the optimized $\alpha$-Te
monolayer structure. The primitive cell is indicated by the blue
shading in the top view. \textit{a} and \textit{b} are the lattice
vectors spanning the 2D lattice. (c)(d) Electronic band structure
and density of states (DOS) of $\alpha$-Te at HSE06 level. The
Fermi levels are set to zero.}
\end{figure*}

\newpage

\begin{figure*}[tbp]
\begin{center}
\includegraphics[width=5.5in]{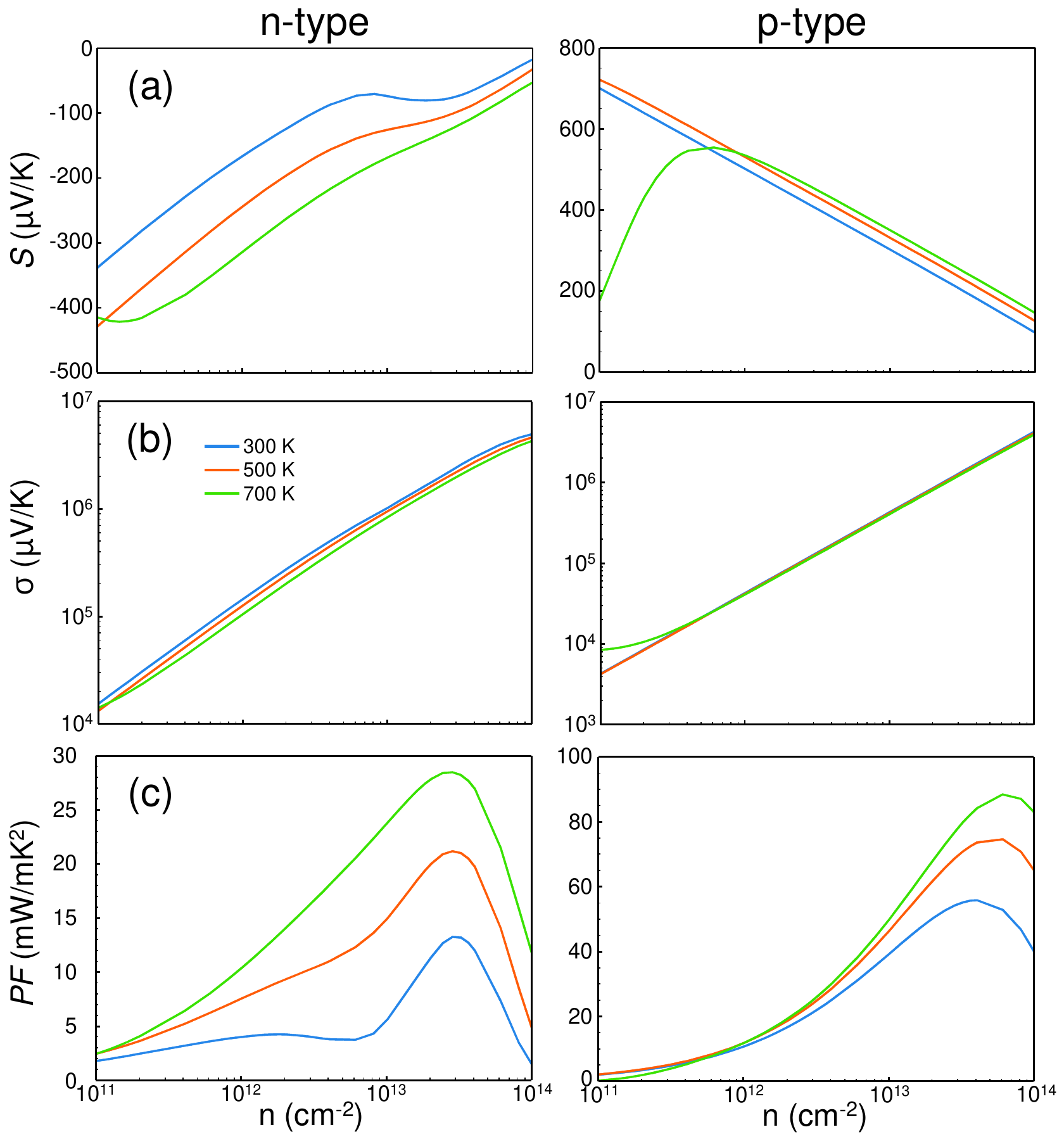}
\end{center}
\caption{Temperature-dependent electronic transport coefficients.
(a) Seebeck coefficient, (b) electrical conductivity, and (c) power
factor (\textit{S}$^2\sigma$) of the $\alpha$-Te as a function of
carrier concentration along \textit{a} or \textit{b} axis at 300,
500, and 700~K.}
\end{figure*}

\newpage

\begin{figure*}[tbp]
\begin{center}
\includegraphics[width=5.5in]{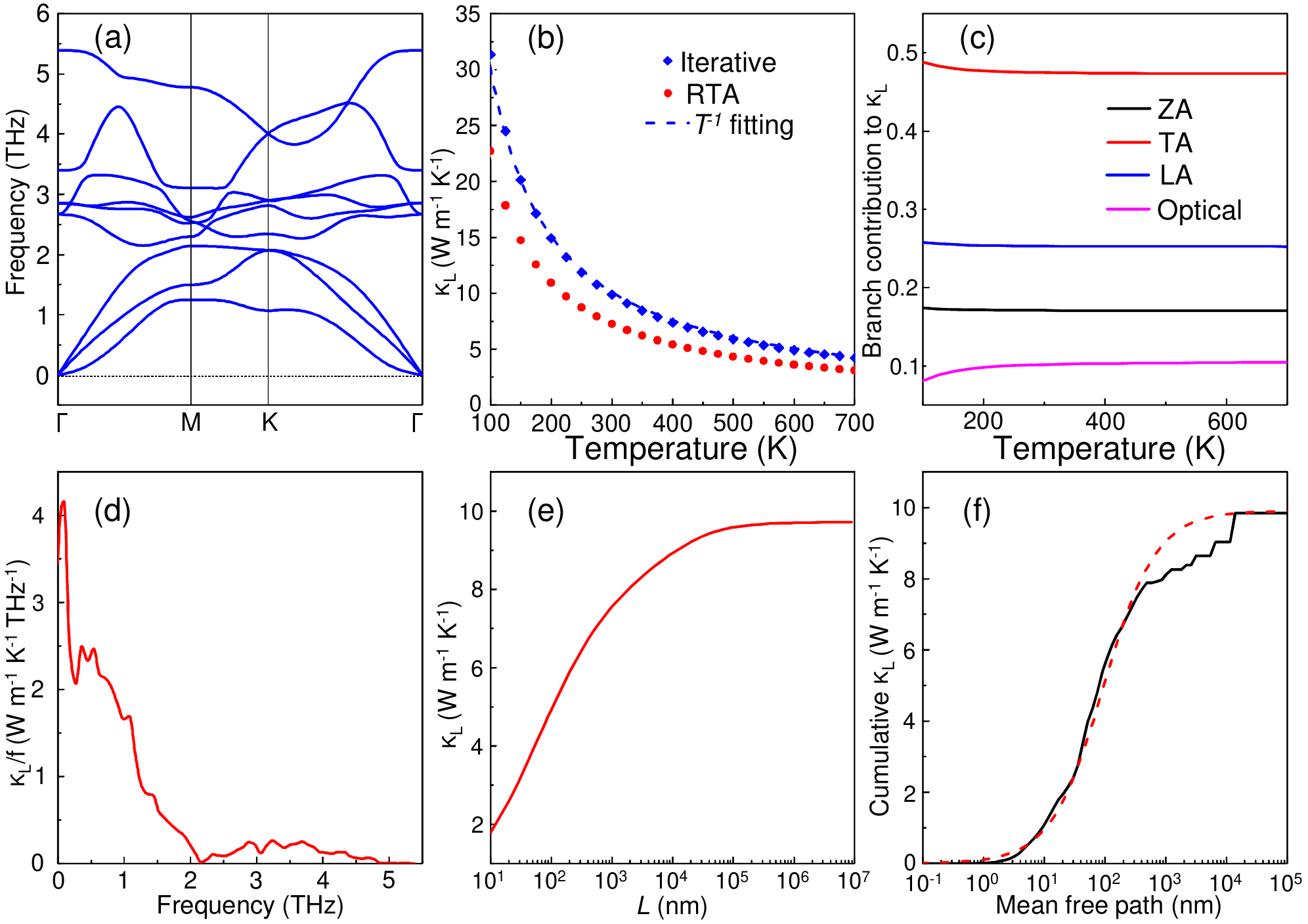}
\end{center}
\caption{(a) Phonon spectra and DOS of $\alpha$-Te. (b) Intrinsic lattice
thermal conductivity $\kappa_L$ as a function of temperature. Blue dashed
line is \textit{T$^{-1}$} fitting. The values of relaxation time
approximation (RTA) are also shown by red dots. (c) Normalized contributions
to total $\kappa_L$ of phonon branches from 100 to 700 K. (d) Frequency-resolved
$\kappa_L$ at room temperature. (e) Room-temperature $\kappa_L$ as a function
of sample size. (f) Cumulative $\kappa_L$ as a function of phonon mean free
path (MFP) at room temperature. The fitted curve is plotted in a dashed line.}
\end{figure*}

\newpage

\begin{figure*}[tbp]
\begin{center}
\includegraphics[width=6.0in]{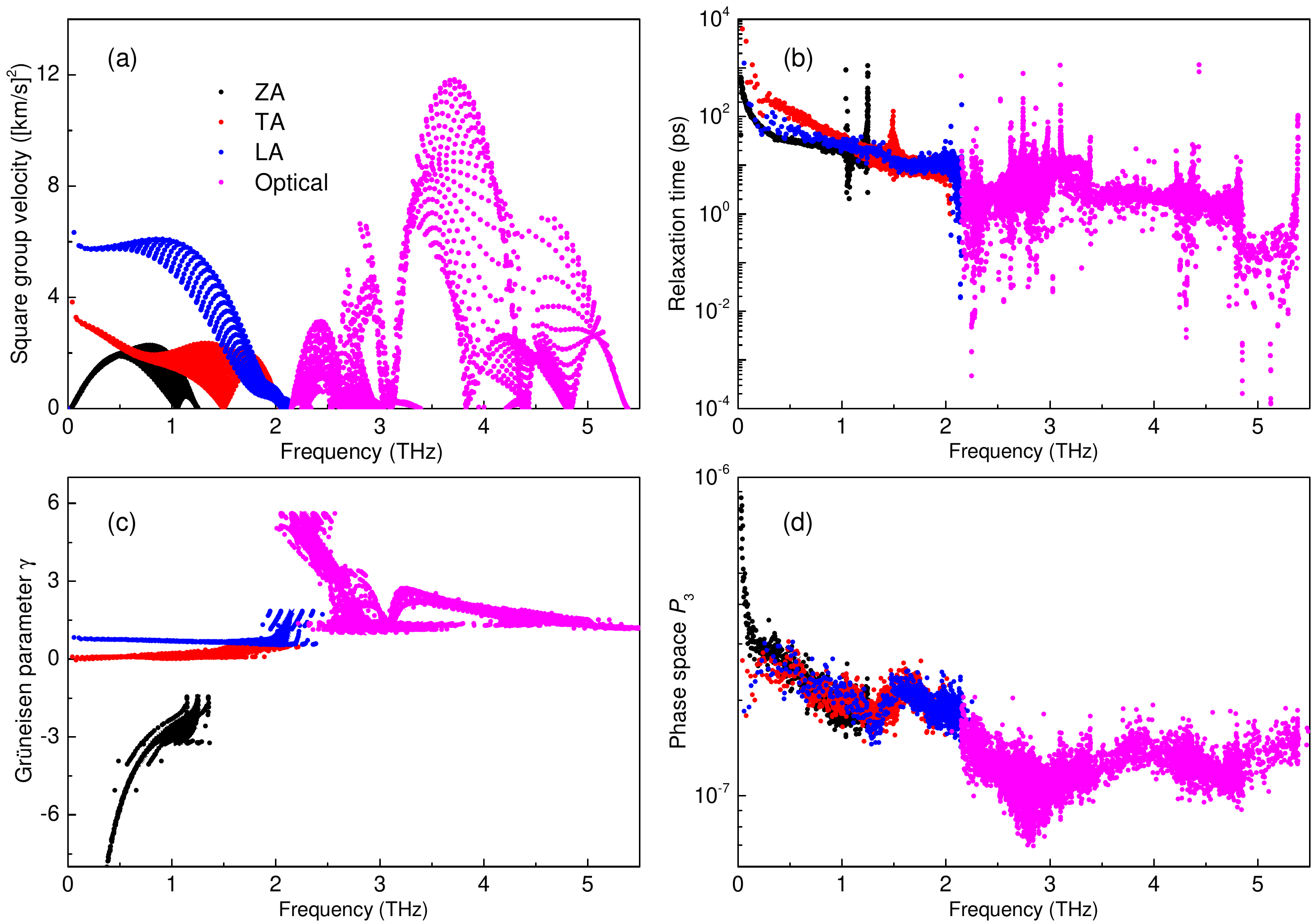}
\end{center}
\caption{(a) The frequency dependent behavior of square group velocity at
300 K for $\alpha$-Te. All optical phonon branches are in purple.
(b) Phonon relaxation time of ZA, TA, LA, and optical phonon branches
as a function of frequency at room temperature. (c) Mode-Gr\"{u}neisen
parameter and (d) Phase-space volume \textit{P$_3$} for three-phonon
scattering processes.}
\end{figure*}

\newpage

\begin{figure*}[tbp]
\begin{center}
\includegraphics[width=6.0in]{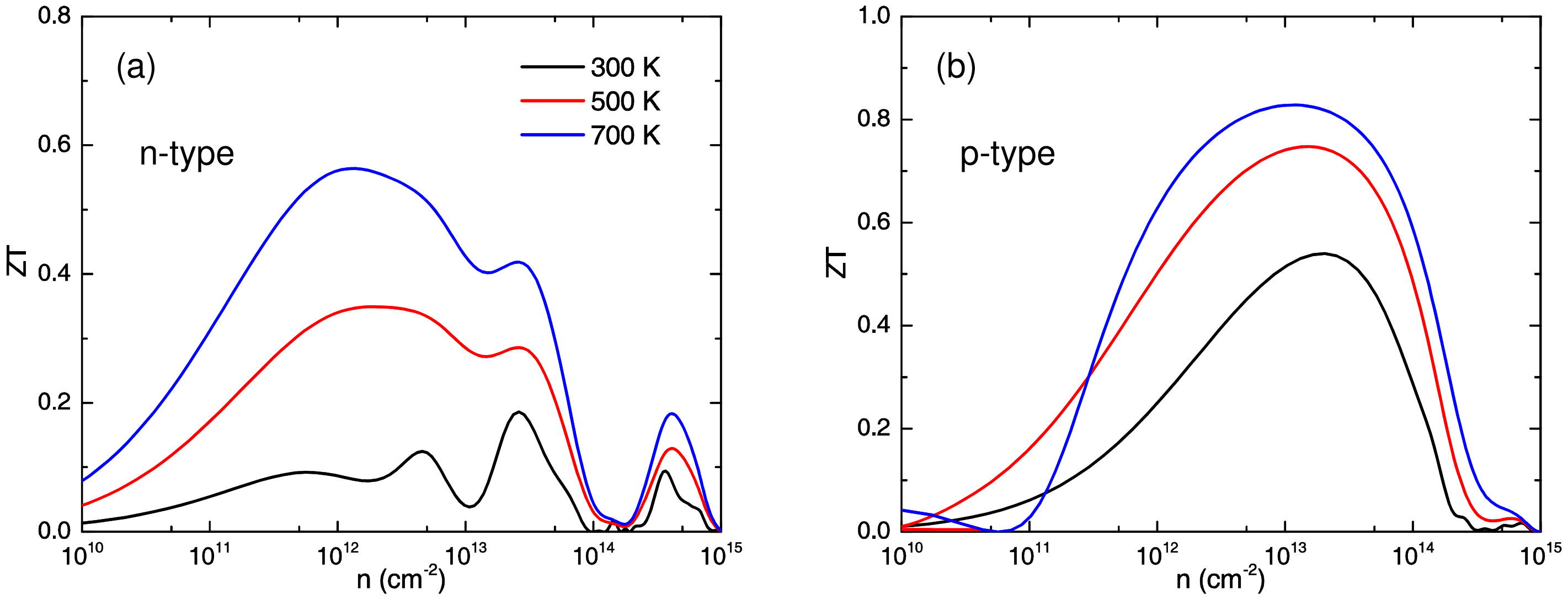}
\end{center}
\caption{Calculated (a) n-type and (b) p-type \textit{ZT} values of
monolayer $\alpha$-Te as a function of carrier concentration along
\textit{a} or \textit{b} axis at 300, 500 and 700~K.}
\end{figure*}

\newpage

\begin{figure*}[tbp]
\begin{center}
\includegraphics[width=6.0in]{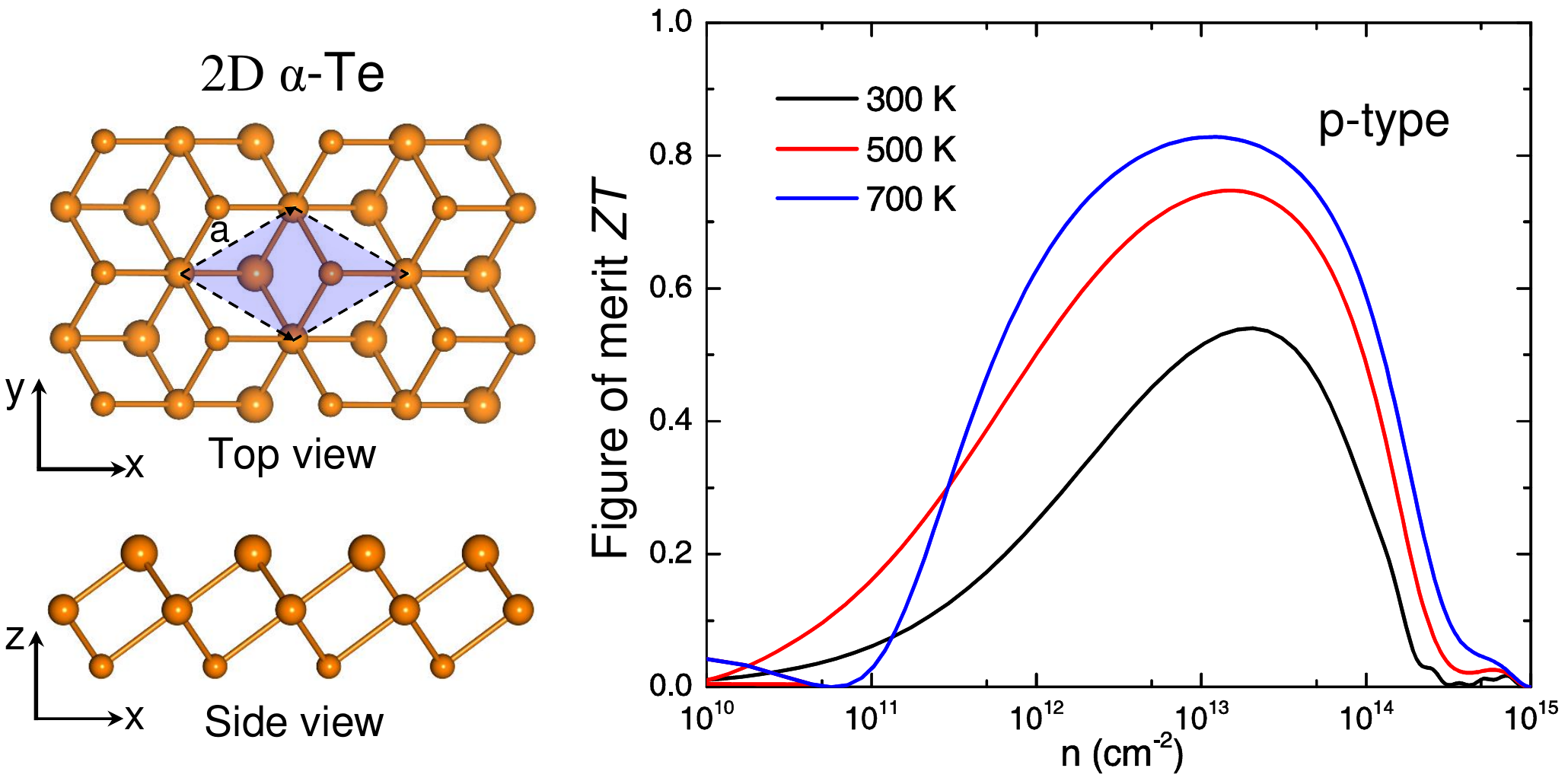}
\captionsetup{labelformat=empty}
\caption{Table of Contents Graphic}
\end{center}
\end{figure*}

\providecommand{\latin}[1]{#1}
\makeatletter
\providecommand{\doi}
  {\begingroup\let\do\@makeother\dospecials
  \catcode`\{=1 \catcode`\}=2 \doi@aux}
\providecommand{\doi@aux}[1]{\endgroup\texttt{#1}}
\makeatother
\providecommand*\mcitethebibliography{\thebibliography}
\csname @ifundefined\endcsname{endmcitethebibliography}
  {\let\endmcitethebibliography\endthebibliography}{}

\end{document}